\documentclass[twoside]{article}
\usepackage{fleqn,espcrc2}
\usepackage{amsmath}
\usepackage{times}
\usepackage{graphicx}
\title{Magnetic  condensation and confinement in lattice gauge theory}
\author{
P. Cea\address{Dipartimento di Fisica - Universit\`a di Bari - 
Via Amendola 173 -
70126 Bari - Italy}$^{\text{,b}}$,
and L. Cosmai\address{INFN - Sezione di Bari - Via Amendola 173 
- 70126 Bari - Italy}
}

\begin{document}

\begin{abstract}
We investigate Abelian monopoles and Abelian vortices  condensation in finite
temperature SU(2) and SU(3) pure lattice gauge theories. 
\end{abstract}
\maketitle

%%%%%%%%%%%%% Introduction %%%%%%%%%%%%
\section{INTRODUCTION}

To study the vacuum structure of lattice gauge theories we
introduced~\cite{Cea:1997ku,Cea:1999gn}
a gauge invariant effective action, defined by using the lattice
Schr\"odinger functional
\begin{equation}
\label{Zetalatt}
{\mathcal{Z}}[U^{\mathrm{ext}}_\mu] = \int {\mathcal{D}}U \; e^{-S_W} \,,
\end{equation}
where $S_W$ is the standard 
Wilson action and the functional
integration is extended over links on a lattice with the
hypertorus geometry and satisfying the 
constraints
\begin{equation}
\label{coldwall}
U_\mu(x)|_{x_4=0} = U^{\mathrm{ext}}_\mu(\vec{x})  \,.
\end{equation}
The same constraint also applies to the links exiting from sites belonging 
to the spatial boundaries (this last condition corresponds to require that the
fluctuations over the background field vanish at infinity).
The lattice effective action 
for the external static background field
$U^{\mathrm{ext}}(\vec{x})$ is
\begin{equation}
\label{Gamma}
\Gamma[U^{\mathrm{ext}}] = -\frac{1}{L_4} \ln \left\{
\frac{{\mathcal{Z}}[U^{\mathrm{ext}}]}{{\mathcal{Z}}(0)} \right\} \,,
\end{equation}
$L_4$ is the extension in Euclidean time, 
$\mathcal{Z}(0)$ is the lattice Schr\"odinger functional,
 without the external background field
($U^{\mathrm{ext}}_\mu = \mathbf{1}$).
It can be shown that: 
 $\Gamma[U^{\mathrm{ext}}]$
is invariant for lattice gauge transformations 
of the external links
$U^{\mathrm{ext}}_\mu$, and
$\lim_{T \to \infty} \Gamma[U^{\mathrm{ext}}] = 
E_0[U^{\mathrm{ext}}] - E_0[0]$, 
$E_0[U^{\mathrm{ext}}]$  vacuum energy in presence 
of the external background field.

At finite temperature,
we consider the 
thermal partition function in presence
of a given static background field:
\begin{equation}
\label{ZetaTnew}
\mathcal{Z}_T \left[ U^{\text{ext}} \right] =
\int_{U_k(\beta_T,\vec{x})=U_k(0,\vec{x})=U^{\text{ext}}_k(\vec{x})}
\mathcal{D}U \, e^{-S_W}   \,.
\end{equation}
Now the temporal links obey p.b.c.'s while spatial links
belonging to the time slice $x_4=0$ or exiting from sites belonging to 
the spatial boundaries are constrained to 
$U_k(\vec{x},x_4)=U_k^{\text{ext}}(\vec{x})$.
After  sending the
physical temperature to zero
the thermal functional
Eq.~(\ref{ZetaTnew}) reduces to the zero-temperature
Schr\"odinger functional with the constraints
$U_k(x)|_{x_4=0} = U^{\mathrm{ext}}_k(\vec{x})$.

To  investigate Abelian
monopoles and Abelian vortices condensation
we make use of a disorder 
parameter~\cite{DelDebbio:1995sx,DiGiacomo:2000fa,Cea:2000zr} 
defined  in terms of the
thermal partition functional Eq.(~\ref{ZetaTnew}) in presence of, respectively,  an
external Abelian monopole or Abelian vortex background field.
Since our thermal partition functional
is constructed by means of the Schr\"odinger functional which is invariant
against gauge transformations of the background field we do not
need to do any gauge fixing to perform the Abelian projection. Indeed, 
after choosing the type of Abelian monopoles, our results 
do not depend on the particular direction selected in the color space, 
which, actually, can be varied by a gauge transformation.

%%%%%%%%%%%%% Abelian monopole condensation: SU(2) %%%%%%%%%%%%
\section{ABELIAN MONOPOLE CONDENSATION}

\subsection{SU(2)}

We put on the background the lattice version of the Abelian monopole field
\begin{equation}
\label{monop3}
g \vec{b}^a({\vec{x}}) = \delta^{a,3} \frac{n_{\mathrm{mon}}}{2}
\frac{ \vec{x} \times \vec{n}}{|\vec{x}|(|\vec{x}| - \vec{x}\cdot\vec{n})} \,,
\end{equation}
$\vec{n}$ is the direction of the Dirac string and, according
to the Dirac quantization condition, $n_{\text{mon}}$ 
is an integer.
The lattice links corresponding to the Abelian monopole field
Eq.~(\ref{monop3}) can be readily obtained as:
\begin{equation}
\label{latlinks}
U^{\text{ext}}_k(\vec{x}) =
\text{P} \exp \left\{
ig \, \int_0^1 dt \,  \frac{\sigma_a}{2}b^a_k(\vec{x} + t \hat{x}_k)
\right\} \,,
\end{equation}
where the $\sigma_a$'s are the Pauli matrices.
By choosing $\vec{n}=\hat{x}_3$ we get:
\begin{equation}
\label{su2links}
\begin{split}
U^{\text{ext}}_{1,2}(\vec{x})  & =
\cos [ \theta_{1,2}(\vec{x}) ] +
i \sigma_3 \sin [ \theta_{1,2}(\vec{x}) ] \,, \\
U^{\text{ext}}_{3}(\vec{x}) & =  {\mathbf 1} \,,
\end{split}
\end{equation}
with
\begin{equation}
\label{thetat3}
\begin{split}
\theta_1(\vec{x}) & = -\frac{n_{\text{mon}}}{4}
\frac{(x_2-X_2)}{|\vec{x}_{\text{mon}}|}
\frac{1}{|\vec{x}_{\text{mon}}| - (x_3-X_3)} \,, \\
\theta_2(\vec{x}) & = +\frac{n_{\text{mon}}}{4}
\frac{(x_1-X_1)}{|\vec{x}_{\text{mon}}|}
\frac{1}{|\vec{x}_{\text{mon}}| - (x_3-X_3)} \,.
\end{split}
\end{equation}
In Equation~(\ref{thetat3}) $(X_1,X_2,X_3)$ 
are the monopole coordinates
and $\vec{x}_{\text{mon}} = (\vec{x} - \vec{X})$.

We evaluate numerically the disorder parameter for confinement
\begin{equation}
\label{disorder}
\mu = e^{-F_{\text{mon}}/T_{\text{phys}}} =
\frac{\mathcal{Z}_T[n_{\text{mon}}]}{\mathcal{Z}_T[0]} \,.
\end{equation}
In Eq.(\ref{disorder}) $\mathcal{Z}_T[0]$ is the thermal partition function
without monopole field (i.e. with $n_{\text{mon}} = 0$).
From Eq.~(\ref{disorder}) it is clear that $F_{\text{mon}}$
is the free energy to create an Abelian monopole.
Therefore in presence of monopole condensation $F_{\text{mon}}=0$
and $\mu = 1$.
Actually we consider the derivative of the monopole free energy
$F^\prime_{\text{mon}} = \partial F_{\text{mon}} / \partial \beta$,
It can be shown that ($V$ is the spatial volume)
\begin{equation}
\label{avplaq}
F^\prime_{\text{mon}} = V \left[ <Pl>_{n_{\text{mon}}=0} -
<Pl>_{n_{\text{mon}} \ne 0} \right] \,.
\end{equation}
$<Pl>$ is the average plaquette.
Figure 1 displays $F^\prime_{\text{mon}}$ vs. $\beta$ together with the absolute value
$|P|$ of the Polyakov loop for a $24^3 \times 4$ lattice. The peak
in   $F^\prime_{\text{mon}}$  corresponds to the rise of $<|P|>$ at the 
deconfinement transition. 
\begin{figure}[t]
\begin{center}
\includegraphics[width=0.48\textwidth,clip]{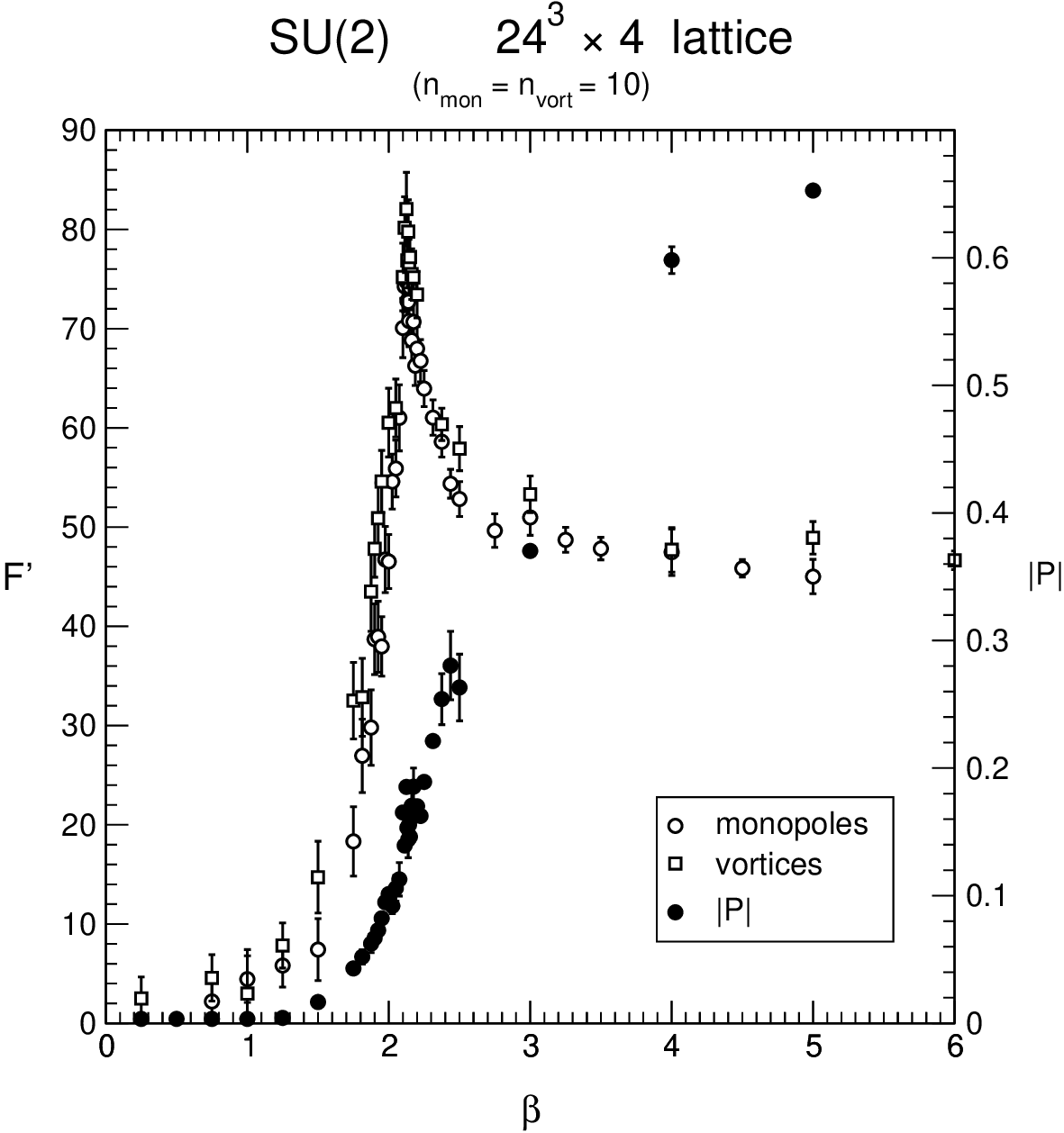}
\caption{$F^\prime$ for monopoles (open circles) and vortices (squares) together
with the absolute value of the Polyakov loop (full circles).}
\end{center}
\vspace{-0.5cm}
\end{figure}
The disorder parameter $\mu$
(Eq.~(\ref{disorder})) can be obtained by a numerical integration in $\beta$ of the
monopole free energy derivative.
It follows that $\mu=1$ in the confined phase (i.e. 
the monopoles condense in the vacuum). Moreover, by increasing the spatial volume 
of the lattice,  the disorder parameter $\mu$ decreases faster toward zero 
in the deconfined phase, suggesting that $\mu \to 0$ in the thermodynamic 
limit when $\beta$ reaches the critical value.

%%%%%%%%%%%%% Abelian monopole condensation: SU(3) %%%%%%%%%%%%
\subsection{SU(3)}

In the case of SU(3) gauge theory, the maximal Abelian group
is U(1)$\times$U(1). We have two different types
of Abelian monopole. 
The first one is defined by Eq.~({\ref{monop3}}) and we call
it $T_3$ Abelian monopole.
The lattice links are given by
\begin{equation}
\label{t3links}
U_{1,2}^{\text{ext}}(\vec{x}) =
\begin{bmatrix}
e^{i \theta_{1,2}(\vec{x})} & 0 & 0 \\
0 &  e^{- i \theta_{1,2}(\vec{x})} & 0 \\
0 & 0 & 1
\end{bmatrix}
\,,
\end{equation}
with $\theta_{1,2}(\vec{x})$ as in Eq.~(\ref{thetat3}).
The second type of independent Abelian monopole can be obtained by
considering the diagonal generator$\lambda_8$: 
\begin{equation}
\label{t8links} 
U_{1,2}^{\text{ext}}(\vec{x}) =
\begin{bmatrix}
e^{i \theta_{1,2}(\vec{x})} & 0 & 0 \\ 0 &  e^{i
\theta_{1,2}(\vec{x})} & 0 \\ 0 & 0 & e^{- 2 i
\theta_{1,2}(\vec{x})}
\end{bmatrix}
\,,
\end{equation}
with
\begin{equation}
\label{thetat8}
\begin{split}
\theta_1(\vec{x}) & = \frac{1}{\sqrt{3}} \left[
 -\frac{n_{\text{mon}}}{4}
\frac{(x_2-X_2)}{|\vec{x}_{\text{mon}}|}
\frac{1}{|\vec{x}_{\text{mon}}| - (x_3-X_3)} \right] \,, \\
\theta_2(\vec{x}) & =  \frac{1}{\sqrt{3}} \left[
+\frac{n_{\text{mon}}}{4} \frac{(x_1-X_1)}{|\vec{x}_{\text{mon}}|}
\frac{1}{|\vec{x}_{\text{mon}}| - (x_3-X_3)} \right] \,.
\end{split}
\end{equation}
We call it the  $T_8$ Abelian monopole.
The corresponding continuum gauge field is now
\begin{equation}
\label{T8monopole} g \vec{b}^a({\vec{x}}) = \delta^{a,8}
\frac{n_{\mathrm{mon}}}{2} \frac{ \vec{x} \times \vec{n}}{|\vec{x}|(|\vec{x}| -
\vec{x}\cdot\vec{n})} \,.
\end{equation}
Other Abelian monopoles can be generated by considering the
linear combination of the  $T_3$ and 
$T_8$ generators. For instance
\begin{equation}
\label{T3a} T_{3a} = -\frac{1}{2} \frac{\lambda_3}{2} +
\frac{\sqrt{3}}{2} \frac{\lambda_8}{2} =
\begin{bmatrix}
0 & 0 & 0 \\ 0 &  \frac{1}{2} & 0 \\ 0 & 0 & -\frac{1}{2}
\end{bmatrix}
\,.
\end{equation}
\begin{figure}[t]
\begin{center}
\includegraphics[width=0.48\textwidth,clip]{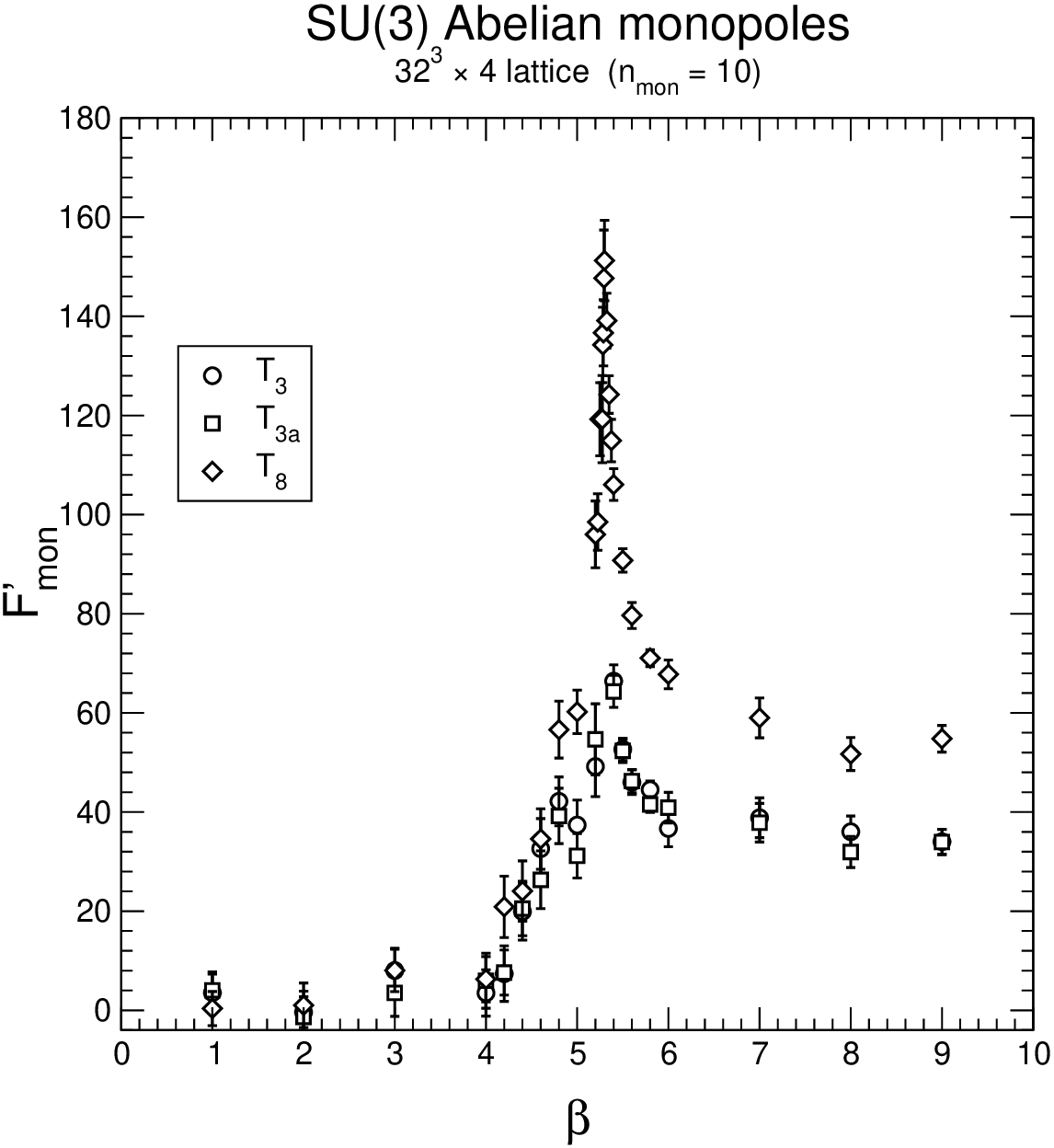}
\caption{$F^\prime_{\text{mon}}$ for $T_3$ monopoles (circles),
$T_{3a}$ monopoles (squares) and $T_8$ monopoles (diamonds).}
\end{center}
\vspace{-0.5cm}
\end{figure}
Figure 2 shows that the $T_8$ Abelian monopole displays a signal about a factor
two higher in the peak region.
This result suggests that in the pattern of dynamical symmetry 
breaking due to the Abelian monopole condensation the 
color direction $8$ is slightly preferred. Like for SU(2),
$F^\prime_{\text{mon}}$ displays a sharp peak in
correspondence of the rise of the Polyakov loop,  and, by increasing the spatial
volume the peak increases.
We find that the disorder parameter $\mu$ is different from
zero in the confined phase and 
decreases towards zero in the
thermodynamic limit at the critical coupling.
Finally a qualitative analysis of $\mu$ 
suggests that the finite
volume behavior is consistent with a first order phase 
transition~\cite{Cea:2000zr}.

%%%%%%%%%%%%% Vortex %%%%%%%%%%%%
\section{VORTEX CONDENSATION}

\subsection{SU(2)}

We consider the continuum gauge field for a vortex directed along the 
$\hat{z}$-direction with $n$ units
of elementary flux $\phi=\frac{2 \pi}{g}$ is given by
\begin{equation}
\label{u1vort}
\begin{split}
A^{\text{ext}}_x &= -\frac{n}{g} \frac{x_2}{(x_1)^2 + (x_2)^2} \,, \\
A^{\text{ext}}_y &=  \frac{n}{g} \frac{x_1}{(x_1)^2 + (x_2)^2} \,, \\
A^{\text{ext}}_z &= A_0 = 0 \,.
\end{split}
\end{equation}
On the lattice 
\begin{equation}
\label{u1vortlinks}
\begin{split}
U^{\text{ext}}_{1,2}(\vec{x})  & =
\cos [ \theta_{1,2}(\vec{x}) ] +
i \sigma_3 \sin [ \theta_{1,2}(\vec{x}) ] \,, \\
U^{\text{ext}}_{3}(\vec{x}) & =  {\mathbf 1} \,,\\
\theta_{1,2}(\vec{x}) & = \mp \frac{n_{\text{vort}}}{2} 
\frac{x_{2,1}}{(x_1)^2+(x_2)^2}
\end{split}
\end{equation}
We evaluate 
$F'_{\text{vort}} = \partial  F_{\text{vort}} \partial \beta$,
where $F_{\text{vort}}$ is the  
free energy to create an Abelian vortex $F_{\text{vort}}$. 
This amounts to compute
\begin{equation}
\label{avplaqvort}
F^\prime_{\text{vort}} = V \left[ <Pl>_{n_{\text{vort}}=0} -
<Pl>_{n_{\text{vort}} \ne 0} \right] \,.
\end{equation}
Figure 1 shows that the free energy  for both monopoles and vortices displays a peak in
correspondence of the finite temperature deconfinement
transition suggesting that vortices play a role in
the dynamics of the deconfinement transition
(in qualitative agreement with Ref.~\cite{DelDebbio:2000cx}).

%%%%%%%%%%%%% Vortex %%%%%%%%%%%%
\subsection{SU(3)}

We have considered
two different kinds of vortices. Namely the $T_8$ vortex:
\begin{equation}
\label{t8vortex} 
\begin{split}
U_{1,2}^{\text{ext}}(\vec{x}) &=
\begin{bmatrix}
e^{i \theta_{1,2}(\vec{x})} & 0 & 0 \\ 0 &  e^{i
\theta_{1,2}(\vec{x})} & 0 \\ 0 & 0 & e^{- 2 i
\theta_{1,2}(\vec{x})}
\end{bmatrix}
\,,\\[0.3cm]
U_3^{\text{ext}}(\vec{x}) &= {\mathbf{1}} \\[0.3cm]
\theta_{1,2} &= \mp \frac{1}{\sqrt{3}} \frac{n_{\text{vort}}}{2} 
\frac{x_{2,1}}{(x_1)^2+(x_2)^2}
\end{split}
\end{equation}
and the $T_3$ vortex:
\begin{equation}
\label{t3vortez}
\begin{split}
U_{1,2}^{\text{ext}}(\vec{x}) &=
\begin{bmatrix}
e^{i \theta_{1,2}(\vec{x})} & 0 & 0 \\
0 &  e^{- i \theta_{1,2}(\vec{x})} & 0 \\
0 & 0 & 1
\end{bmatrix}
\,,\\[0.3cm]
U_3^{\text{ext}}(\vec{x}) &= {\mathbf{1}} \\[0.3cm]
\theta_{1,2} &= \mp  \frac{n_{\text{vort}}}{2} 
\frac{x_{2,1}}{(x_1)^2+(x_2)^2}
\end{split}
\end{equation}
\begin{figure}[t]
\begin{center}
\includegraphics[width=0.48\textwidth,clip]{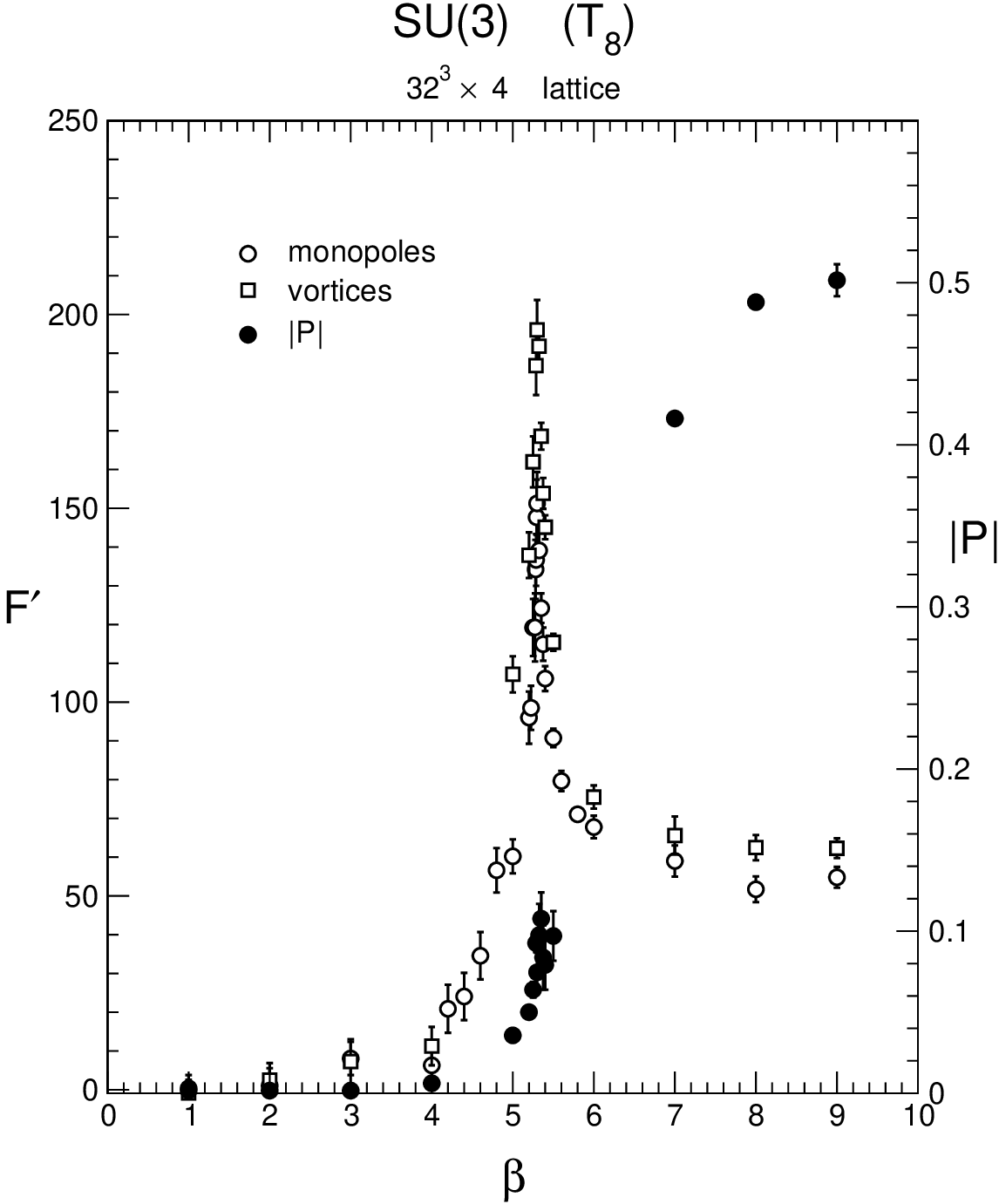}
\caption{$F^\prime$ for monopoles (open circles) and vortices (squares) together
with the absolute value of the Polyakov loop (full circles).}
\end{center}
\vspace{-0.5cm}
\end{figure}
Also in this case we clearly see (Fig. 3)
a peak in $F'_{\text{vort}}$ corresponding to the rise of the absolute
value of the Polyakov loop at the deconfinement transition.
As in the case of Abelian monopoles we find that $F'_{\text{vort}}$ for
the $T_8$ vortex gives a stronger signal than the corresponding quantity
for the $T_3$ vortex.

%%%%%%%%%%%%% CONCLUSIONS %%%%%%%%%%%%
\section{CONCLUSIONS}

We investigated the condensation of Abelian monopoles and Abelian vortices
in the finite temperature SU(2) and SU(3) lattice
gauge theories using a disorder parameter defined in terms of
a lattice thermal partition
functional.
Our numerical results suggest that monopoles and vortices condense
in the vacuum of non Abelian l.g.t.'s., in agreement with
analogous results obtained using different approaches~\cite{Topology}.

Let us conclude by stressing that our method,  while keeping the gauge
invariance, can be readily extended to study the effect of the dynamical
fermions.

%%%%%%%%% Bibliography  %%%%%%%%%%%%%%%%%%%%%%%%%%%%%%%%%%%%%%%%%%%%%%%%%%%%%%%%%%%%

\end{document}